\documentclass{aastex}
\usepackage{emulateapj5}
\received{2002 August 2}
\begin{document}
\newcommand{\Halpha}{H$\alpha$ }
\newcommand{\etal}{\mbox{et al.}}
\newcommand{\NHoo}{N_{\rm H}}
\newcommand{\NHxo}{N_{\rm x}}
\newcommand{\NHIo}{N_{\rm HI}}
\newcommand{\NHtwoone}{N$_{\rm 21cm}$ }
\newcommand{\NHII}{N_{\rm HII}}
\newcommand{\NHgx}{N_{\rm G}}
\newcommand{\acm}{ cm$^{-2}$ }
\newcommand{\as}{s$^{-1}$}
\newcommand{\Htwo}{H$_{2}$}
\newcommand{\LHB}{{\sc lhb}}
\newcommand{\ISM}{{\sc ism}}
\newcommand{\LISM}{{\sc lism}}
\newcommand{\MER}{{\sc mer}}
\newcommand{\ASCA}{{\it ASCA}}
\newcommand{\IRAS}{{\it IRAS}}
\newcommand{\ROSAT}{{\it ROSAT}}
\newcommand{\PSPC}{{\it PSPC}}
\newcommand{\COBE}{{\it COBE}}
\newcommand{\DIRBE}{{\it DIRBE}}
\newcommand{\EUVE}{{\it EUVE}}
\newcommand{\tenup}[1]{\times 10^{#1}}
\newcommand{\gte}{$\infty$\phn}
\newcommand{\ZY}{0.3,1.0}
\newcommand{\ZZ}{0.5,1.0}
\newcommand{\Msun}{$M_{\odot}$}
\newcommand{\ayr}{y$^{-1}$}
\newcommand{\dgr}{$^{\circ}$}
\newcommand{\gp}{\hspace*{25pt}}
\newcommand{\gq}{\hspace*{19pt}}

\title{Small-Scale structure in the Galactic ISM: \\
Implications for Galaxy Cluster Studies}

\author{Joel N.\ Bregman\altaffilmark{1}, Megan C.\ Novicki\altaffilmark{2},
Jessica E.\ Krick\altaffilmark{1}, and John S.\ Arabadjis\altaffilmark{3}}
\altaffiltext{1}{University of Michigan, Department of Astronomy, 833 Dennison
Building, Ann Arbor, MI 48109}
\altaffiltext{2}{Institute for Astronomy, University of Hawaii, 2680 Woodlawn
Drive, Honolulu, HI 96822}
\altaffiltext{3}{Massachusetts Institute of Technology, Center for Space
Research, 70 Vassar Street, Room 37-501, Cambridge, MA 02139}

\begin{abstract} 

Observations of extragalactic objects need to be corrected for Galactic
absorption and this is often accomplished by using the measured 21 cm HI
column.  However, within the beam of the radio telescope there are
variations in the HI column that can have important effects in interpreting
absorption line studies and X-ray spectra at the softest energies.  We
examine the HI and DIRBE/IRAS data for lines of sight out of the Galaxy,
which show evidence for HI variations in of up to a factor of three in
1\dgr\ fields.  Column density enhancements would preferentially absorb soft
X-rays in spatially extended objects and we find evidence for this effect in
the \ROSAT\ \PSPC\ observations of two bright clusters of galaxies,
Abell 119 and Abell 2142.

For clusters of galaxies, the failure to include column density fluctuations
will lead to systematically incorrect fits to the X-ray data in the sense
that there will appear to be a very soft X-ray excess.  This may be one
cause of the soft X-ray excess in clusters, since the magnitude of the
effect is comparable to the observed values.\bigskip 

\end{abstract}


\section{Introduction} 

In the course of analyzing X-ray observations of extragalactic objects,
corrections are applied for the effects of Galactic absorption.  The amount
of Galactic absorption can be fit directly from the observations for
adequately strong sources with a known underlying spectrum, and this value
can be compared to the amount of HI measured from 21 cm line emission.  The
similarity or difference between these two measures of the absorption column
can be used for a variety of purposes, such as to examine whether there is
excess absorption within a cluster of galaxies \citep{Wetal,AB00}
or whether absorption by molecular material is commonplace
within the Galaxy \citep{AB99a}.  In other cases, one can
fix the X-ray absorption column at the 21 cm HI value and examine whether
there is an additional emission component in a system, such as the very soft
X-ray excess (the EUVE excess) that has been claimed to exist in clusters of
galaxies (e.g., \citet{LBM}; but see \citet{BBK} and \citet{AB99b}).

In these studies, one usually ignores the small-scale structure of the
neutral Galactic ISM, assuming that the mean 21 cm column density within a
radio telescope beam is uniform across the face of the beam, which is
typically 30$^{\prime }$ for the single-dish radio telescopes that most of
the HI values are based upon.  Naturally, there is structure to the total
neutral column within the radio telescope beam and this has the potential of
influencing the analysis of extragalactic X-ray emission as well as
absorption studies in the optical and ultraviolet region. Here we examine
the magnitude of the variation of the small-scale structure in the neutral
Galactic layer and we discuss the implications for the analysis of data.
First, we present an idealized model to illustrate the magnitude of the
effect (\S 2), and then we present the HI and X-ray observations that show
the magnitude of the variation in the absorption column (\S 3, 4).  Among
other effects, we show that by ignoring such structure in the ISM,
absorption corrected X-ray spectra of clusters will systematically show an
apparent excess soft emission component, of the sort that is seen in some
EUVE observations of galaxy clusters (\S 5).

\section{An Example of the Importance of Small-Scale Structure} 

The importance of small-scale structure on absorption corrections can be
illustrated by adopting simple models for the fluctuations in $N_{H}$ that
are sensible and are consistent with the observations discussed below.  In
one model, we assume that the column density distribution $f(N)$ has a
minimum level, $N_{low}$, a maximum level, $N_{high}$, and behaves as a
power-law between these levels so that
\begin{eqnarray}
 f(N) = & N^{-m} & \gp {\rm for} \gp N_{low} \leq N\leq N_{high} \nonumber \\
 f(N) = & 0      & \gp {\rm for} \gp N<N_{low}, \ N_{high}<N
\label{eq01}
\end{eqnarray}

The mean column density is the 21 cm HI\ column, which is

\begin{eqnarray}
N_{avg} = N_{21cm} & = &
\frac{\int_{N_{low}}^{N_{high}}fNdN}{\int_{N_{low}}^{N_{high}}fdN} \nonumber \\
\phn{ } & = & \frac{m-1}{m-2} \,
\left(
\frac{N_{low}^{-(m-2)} - N_{high}^{-(m-2)}}{N_{low}^{-(m-1)}-N_{high}^{-(m-1)}} \right),
\end{eqnarray}

\noindent and in the case of $N_{high}=\infty $, we have the simple form
$N_{21cm}=\frac{m-1}{m-2}N_{low}$.  For individual lines of sight out of the
Galaxy, one is more likely to measure a column that is typically closer to
the median column density rather than the average column density $N_{avg}$.
The median column density, or the column density most often be sampled, is
\begin{equation}
N_{median}=2^{1/(m-1)} \, (N_{low}^{-(m-1)}+N_{high}^{-(m-1)})^{-1/(m-1)}
\end{equation}

\noindent and in the limit of $N_{high}=\infty $,
$N_{median}=2^{1/(m-1)}N_{low}$.  Then, the ratio of the median to the mean
is, in the limit of $N_{high}=\infty $,
\begin{equation}
N_{\rm median}/N_{21cm} = \frac{(m-2) \, 2^{1/(m-1)}}{m-1}.
\end{equation}

If the distribution function were infinitely sharply peaked,
$N_{median}=N_{21cm}$, but if the distribution has some significant width,
these two quantities can be measurably different.  For example, with a
value of $m=3.5$, consistent with the data of \citet{WOP},
$N_{median}/N_{21cm}=0.792$, or about a 20\% difference between the
median and the mean.  This implies that individual lines of sight out of
the Galaxy will generally return a value for the column that is about 20\%
lower than the column induced from the 21 cm measurements.  As an example
of the importance of this in an astrophysical setting, this phenomenon will
affect the interpretation of absorption line measurements (single lines of
sight) when they are compared to the amount of 21 cm emission (as was first
pointed out to us by Wakker).  In a comparison of a Ly$\alpha $ absorption
line to a 21 cm emission line measurement, one would find the systematic
effect that the absorption column was less than the emission column.  In
the determination of metallicities that compare metal absorption lines to 21
cm emission, one would determine a lower than true metallicity more often
than a higher metallicity.

These small scale variations in the HI column will affect the absorption of
light in the X-ray waveband in significant ways, especially at high optical
depths.  The transmitted fraction of light at some energy is given by
\begin{equation}
Tr(E) = I_{o}(E) \, \frac{\int e^{-\sigma (E)N(x,y)} \, dx \, dy}{\int dx \, dy}
\end{equation}

\begin{figure}
\epsscale{0.8}
\plotone{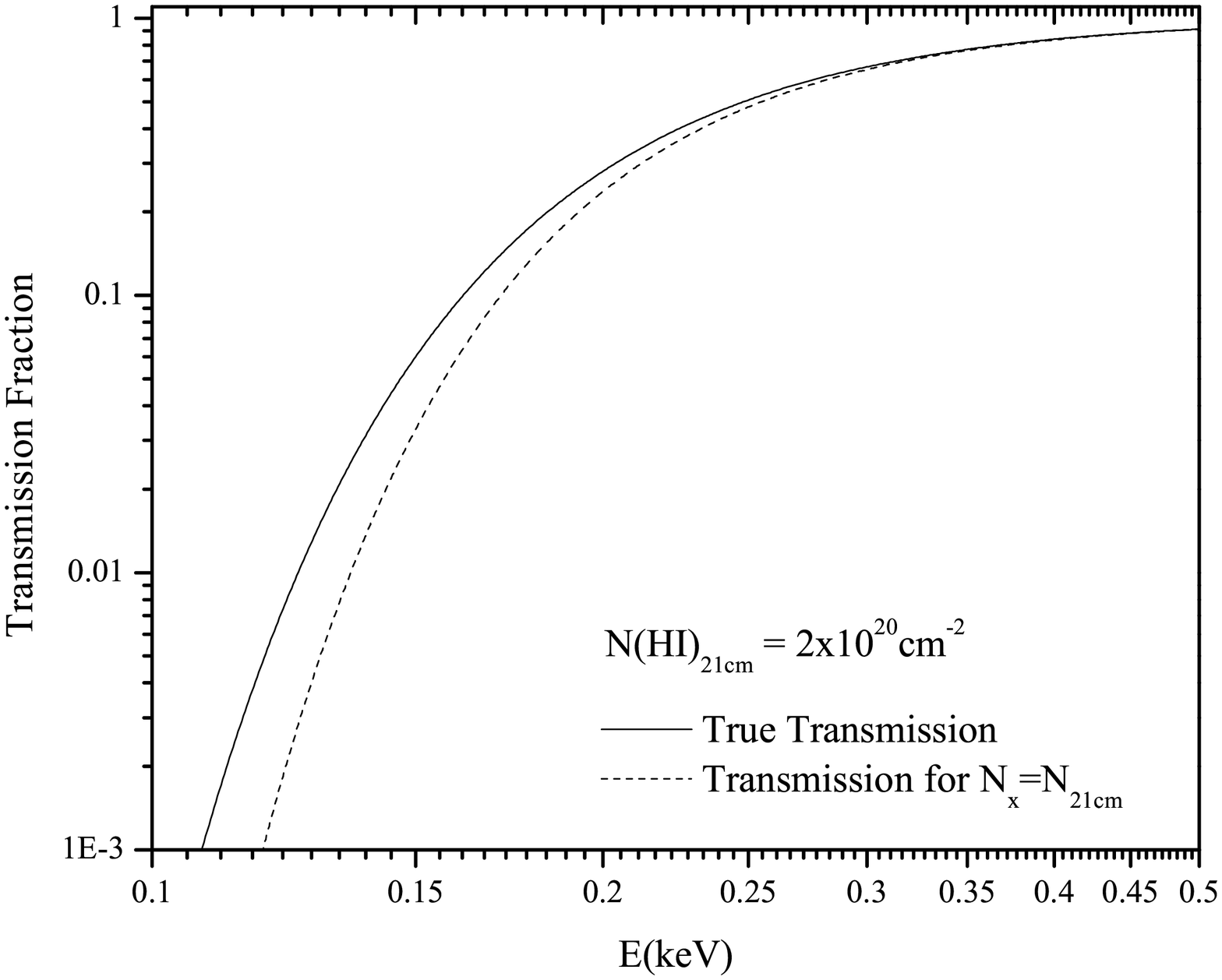}
\caption{The transmission fraction due to neutral atomic gas for a column
of $2\tenup{20}$\acm\ (dashed line) compared to the transmission for
gas of the same mean column but with fluctuations as described in the text.
The amount of light transmitted through the absorbing medium is less for a
uniform absorption column then one that has fluctuations, so the failure to
include such fluctuations will lead to a systematic residual at energies
where the optical depth exceeds unity.
\label{f01}}
\end{figure}

\noindent At low optical depth ($\tau (x,y,E)=\sigma (E)N(x,y)<<1$, 
where $\sigma (E)$ is the energy dependent cross section), this
just reduces to $Tr(E)=I_{o}(E)(1-\sigma (E)N_{avg})$, so it is appropriate
to use $N_{avg}=N_{21cm}$, as is usually done.  However, if there are
variations in $N(x,y)$ when $\tau >1$, the effective X-ray absorption column
changes with energy, with the deviation from $N_{21cm}$ becoming greater
toward lower energies, which is the direction of increasing optical depth.
This can be illustrated in Figure~\ref{f01}, where we show the transmitted
fractions as a function of energy for the case where
$N_{avg} = N_{21cm} = 2\tenup{20}$\acm, and where the variation occurs from
$N_{low} = 0.82 \, N_{avg}$ to $N_{high} = 1.31 \, N_{avg}$, with $m = 3.5$.
At high optical depths, the transmission is underestimated when using
$N_{21cm}$.  The reason is due to the non-linearity in the behavior of the
transmission with column, so that the fluctuations toward low density (lower
optical depth) permit far more photons to flow through than an equivalent high
density fluctuation removes.  This leads to a residual flux when fitting a
model with constant $N_{21cm}$ to the data, and this positive residual is a
systematic effect that is prominent in a moderately narrow energy range where
$4\gtrsim \tau \gtrsim 1$.  The absolute value of the residual is small at low
energies because the optical depth through the Galaxy is large, while the
residual vanishes at larger energies because the optical depth becomes low
enough that using a constant $N_{21cm}$ is a good approximation.  This effect
can be rather important in the discussion of the reality of the soft excess
X-ray emission, as we discuss below.  However, first it is important to
understand the magnitude of the fluctuations in $N$ so that the importance
of the effects can be assessed.

\section{The Magnitude of Small-Scale Structure in the ISM from 21 cm Studies}

Most extragalactic objects that have been studied lie at Galactic latitudes 
$b > 20$\dgr, or lines of sight that are out of the plane of the Milky Way.
For such lines of sight, there are few studies of the small-scale structure in
the total HI column, and almost no information on the structure of H$_{2}$.
The type of information needed is the distribution function of column
densities within the larger beam of single-dish surveys (i.e., a histogram
of N$_{HI}$).  In principle, this can be obtained with synthesis arrays,
but such instruments normally resolve away the larger scale structure,
showing only small-scale variations.  However, synthesis array
observations, used in conjunction with single-dish observations (the zero
spacing data) can recover all the emission and show the true distribution of
N$_{HI}$.  Unfortunately, few such observations have been made out of the
plane, although there are many such observations in the plane, which show a
rich amount of structure.  Out of the plane, two published observations are
by \citet{WOP}, which is toward a high velocity cloud (WW 187) in the direction
of NGC 3783, $l=287$\dgr, $b=23$\dgr, and by \citet{JBD}, which is toward a
filament of far infrared cirrus emission at $l=141$\dgr\, $b=30$\dgr.  Although
these may not be typical regions, we examine them since they provide a starting
point for understanding small-scale structure; we will be obtaining 21 cm data
toward more typical sight lines in the near future.

\begin{figure}
\epsscale{0.8}
\plotone{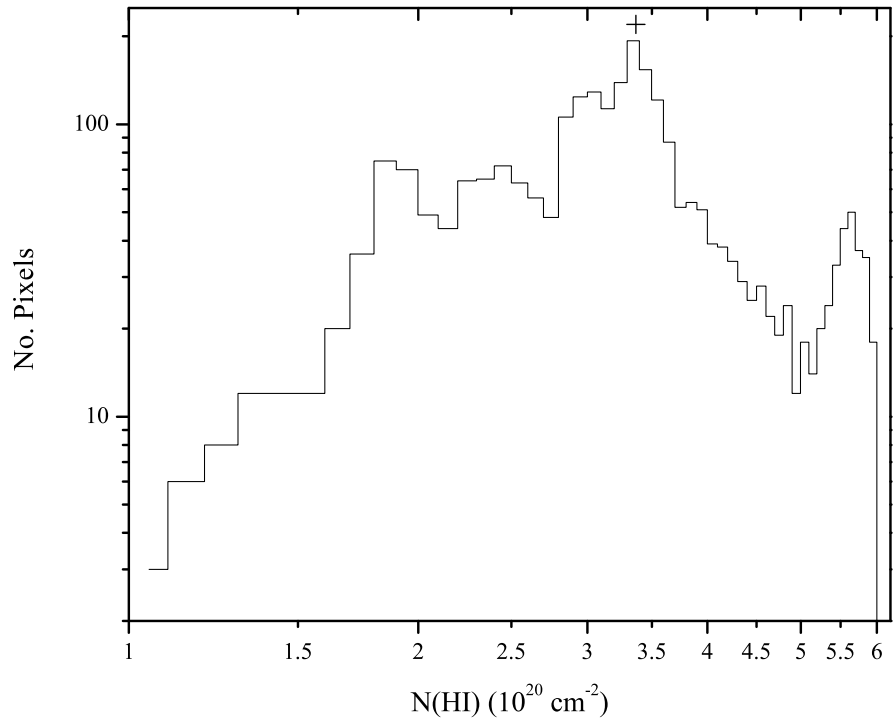}
\caption{Histogram of the HI column density in the central 1\dgr\
field that was obtained by Joncas \etal\ (1992).  The few very low column
density pixels is due to the S/N of the observation (an rms of 5 K).
The range in the column density is more than a factor of three in this region, 
which includes a filament identified in \IRAS\ data.
\label{f02}}
\end{figure}

The observation by \citet{JBD} is toward a gas and dust complex in Ursa
Major that has conspicuous IR-cirrus emission and is also known to have some
molecular gas.  However, the mean HI column in this direction is
3.4$\tenup{20}$\acm, which is fairly typical for sight lines at
this Galactic latitude.  The observation was made with the Synthesis Radio
Telescope of Dominion Radio Astrophysical Observatory, including zero
spacing data and complementary observations in the continuum; the resulting
resolution was 1$^{\prime }$ and the field of view was 2.6\dgr.  Since the
sensitivity is dropping toward the edges, and the absolute calibration
suffers, we have chosen to look at the variations in the central 1\dgr\
region, smoothed to 2.3$^{\prime }$ and formed a histogram of the column
densities (Figure~\ref{f02}).  There is a fairly sharp cutoff in N(HI) for
values below 1.7$\tenup{20}$\acm, but there is a distribution that extends to
$5.9\tenup{20}$\acm, a range of almost a factor of four (0.54 in the log).

\begin{figure}
\epsscale{0.8}
\plotone{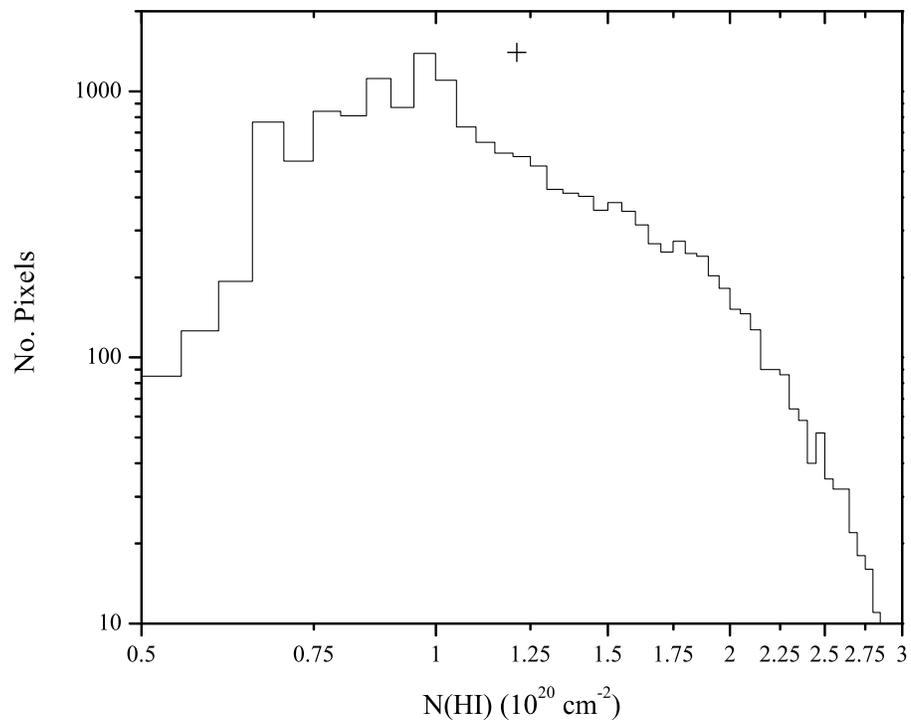}
\caption{Histogram of the HI column density in the field around WW 187
obtained by \citet{WOP}.  A power-law of slope -3.5 would be steeper than the
observed distribution for
$0.8\tenup{20} {\rm cm}^{-2} < N_{21cm} < 2\tenup{20} {\rm cm}^{-2}$,
and it would be shallower than the observed distribution at greater columns.
\label{f03}}
\end{figure}

The observation by Wakker is of a HVC projected upon a background AGN (NGC\
3783) in the Southern Hemisphere, for the purpose of understanding the
optical and ultraviolet absorption line results toward NGC 3783.  The
synthesis observations were made with the Australia Telescope Compact Array,
supplemented with data from the Parkes Multibeam survey, so that HI\
structures were not resolved out of the image.  The map that they produced
was not of all of the Galactic HI in that region, but only the HI in the
velocity range surrounding cloud, or 190-280 km s$^{-1}$.  Although it
would be better to have the entire Galactic column, such a map was not
produced, so we have analyzed the existing map, which should give us some
insight into HI variations.  We have used a square box 1\dgr\
on a side located in the central region in order to determine the column
density histogram (Figure~\ref{f03}), which shows that there is a sharp drop in
the column below $0.65\tenup{20}$\acm.  The HI distribution is
moderately flat to $1.2\tenup{20}$\acm, declining toward higher
values ($16\tenup{20}$\acm\ in a way that may be described as a
broken power-law of slopes -3 to -4 (or with an exponential model).  A
detailed analysis of this distribution is given in \citet{WOP}.

\begin{figure}
\epsscale{0.8}
\plotone{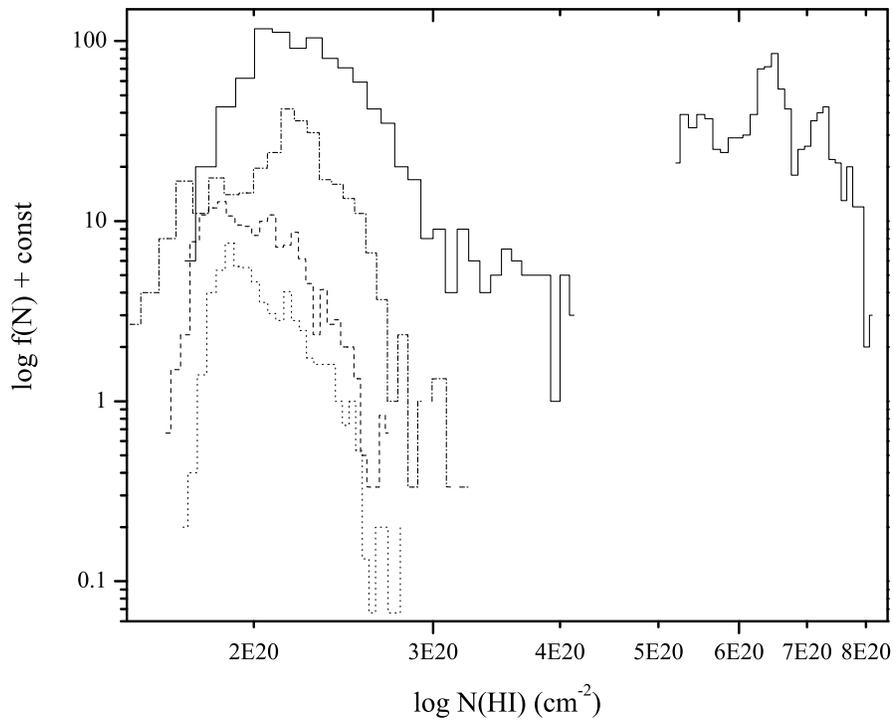}
\caption{The column density distribution within 1\dgr\ fields inferred from
the IRAS-DIRBE data of \citet{SFD}; arbitrary vertical shifts are introduced to
separate the histograms.  In general, there is a peak in the distribution, with
a greater extension to higher column than to lower columns, and in some cases,
there is a sharp cutoff at the low column density side.
\label{f04}}
\end{figure}

Although there are few HI maps such as the above, the entire sky has been
mapped in the far infrared with DIRBE and IRAS data by \citet{SFD}.  These
authors have produced maps of E(B-V), which are, on average, proportional to
the HI column density, and with 2.67$^{\prime }$ resolution.  
We sampled five randomly chosen 1\dgr\ square regions
located at $l,b$ = $-$141.3\dgr, 37.3\dgr; 14.0\dgr, 74.8\dgr; 112.8\dgr,
47.7\dgr; $-$44.3\dgr 45.4\dgr; and 52.4\dgr, 55.9\dgr, and analyzed the column
density distribution, using the relationship of
$N(HI) = 8.18\tenup{21} E(B-V) + 2.54\tenup{19}$\acm, which
was determined by us using other lines of sight out of the Galaxy.  It is
evident that the full range is at least 100\% relative to the minimum value
(Figure~\ref{f04}).  However, we can better characterize the range by using the
quartile points, referenced to the median column density, which in
percentage, is $100 \, (N(75\%)-N(25\%))/N_{median}$, where $N(75\%)$ and
$N(25\%)$ are the columns at the 75\% and 25\% quartile locations.  Using this
measure, we find that for these selected regions, the range is fairly
narrow, 14.6-19.2\% with an average value of 16.5\%.

Unfortunately, it is nearly impossible to determine whether the degree of
fluctuations seen in the HI is the same as in the IRAS-DIRBE data at a
resolution of 1$^{\prime }$.  The HI data from Wakker show a greater degree
of HI fluctuations than does the IRAS-DIRBE data, but this may be because
the HI map does not include HI at all velocities.  In the region studied by
Joncas, the HI\ and IRAS-DIRBE variations are similar in magnitude, but this
region was chosen because of the prominent IRAS structure in the field,
which guarantees substantial variation.  One might expect the IRAS-DIRBE
variations to be smaller than the HI variations just because the "beam" of
the IRAS-DIRBE data is an order of magnitude larger than the HI radio
synthesis beam of about 1$^{\prime }$ (so small scale variations are washed
out).  This is an issue where more data would be extremely valuable.

There is additional evidence for small-scale structure in the interstellar
medium, such as from the studies that are based upon optical and ultraviolet
absorption measurements.  The studies of NaI absorption systems toward
globular cluster stars show that there are variations in the equivalent
width by a factor of a few on scales of 1-4$^{\prime }$ \citep{LPS} as well as
on scales of $\sim10^{\prime \prime }$ \citep{AML}.  Absorption line
studies toward binary stars show that there is significant structure in the ISM
on a scale of 0.1$^{\prime \prime }$ (Laroesch, Meyer, and Blades 1999).  In
addition, HI absorption studies of the 21 cm line show that 10-15\% of cold HI
occurs on scales below 0.1$^{\prime \prime }$.  \citet{FWetal} study
the changes in 21 cm absorption toward pulsars with known velocities,
detecting changes in the absorption line over a timescale of a few years,
which corresponds to a change of 10-100 mas.  On comparable angular scales,
\citet{FGetal} used VLBI studies toward radio AGNs and find
variations in the HI optical depth as large as a factor of two.
Unfortunately, neither the optical absorption lines (NaI, KI) or the 21 cm
absorption line is a true linear indicator of the gas column because the
optical metal lines can be sensitive to changes in the ionization state
while the 21 cm optical depth is inversely proportional to the temperature.
\ Nevertheless, these studies are consistent with the presence of structure
in the interstellar column on relatively small scales.

\section{Small-Scale Variation From X-Ray Observations} 

The absorption of soft X-rays depends upon the column of intervening gas as
well as the energy of the observation, since the approximate relationship
for the optical depth is $\tau \varpropto NE^{-3}$, where $N$ is the column
of X-ray absorbing gas (usually expressed as an equivalent HI column) and $E$
is the energy of the photons.  For a column density of $N=2\tenup{20}$\acm,
$\tau =1$ at 0.24 keV, while for a column of $4\tenup{20}$\acm,
$\tau =1$ occurs at 0.32 keV, so if there are significant fluctuations in the
X-ray absorption column, we should see a ``drop-out'' of the soft emission
(the ratio of soft to hard emission decreases).  At these column densities,
the absorption of X-rays is due primarily to HeI, with HI being the second
most important absorber.  Other components of the ISM are less important as
absorbers (warm ionized gas or hot X-ray emitting gas), so the X-ray column
is a fairly good measure of the neutral column.

This effect is, in principle, possible to detect against a relatively smooth
extended extragalactic X-ray emitting object such as a cluster of galaxies.
\ In practice, it is possible to search for this effect only in clusters of
high surface brightness and for the correct range of Galactic column
density, at least when using ROSAT data.  The ROSAT PSPC data has rather
limited energy resolution so that the data can only be separated into a two
or three independent spectral regions.  One of these divisions is due to
the presence of carbon in the optical path, which leads to a sharp
absorption edge at 0.284 keV that becomes progressively less important at
higher energies.  Operationally, this leads to a local minimum in the
detection efficiency near 0.5 keV, so we have divided the PSPC data into two
bands, 0.2-0.5 keV and 0.5-2.0 keV (PI channels 20-51 and 52-201).  Snowden
\etal\ (1994) have developed procedures to extract diffuse emission in
various energy bands, corrected for point sources, instrumental background,
and local background sources (solar scattered X-rays, etc.).  Here we use
those procedures to form the ratio of band R2+R3 (the 0.2-0.5 keV band) to
R4+R5+R6+R7 (0.5-2.0 keV).

In order to detect variations in the ratio of these two bands, there must be
sufficient photons in the soft band for statistical reasons and that implies
a mean Galactic absorption column of less than about $6\tenup{20}$\acm\
(otherwise there are no soft photons) but the absorption in the soft
band should be significant ($\tau =1$) so that modest variations in $\tau $
lead to detectable variations in the X-ray ratio.  For example, if the mean
Galactic column is $3\tenup{20}$\acm\ and there is a region where the column
rises to $4\tenup{20}$\acm, our hardness ratio will decrease by 30\% while if
there is a region where the column rises to $5\tenup{20}$\acm, the hardness
ratio will drop to half its original value.  From the above discussion,
variations of this sort are not likely, although the probabilities of finding
such variations can depend strongly on region.  For the HI data of \citet{JBD}
and \citet{WOP} about 20\% of the data are above the mean by 4/3 and
10\% of the data are above the mean by a factor of 5/3 or greater.  However,
if one uses the variations in the DIRBE/IRAS data for these two regions and
the five other regions (above), typically only 1\% of the area has an
intensity value that is above the mean by 4/3 or greater.  Since the variation
in the DIRBE/IRAS data is less than that in the HI for the region observed by
\citet{WOP}, this latter value of 1\% may be an underestimate.  This suggests
that finding significant variation in our hardness ratio is likely if there
are several hundred independent positions.

We analyzed five clusters as given in Table~\ref{t01}, all of which are bright
nearby systems that are fairly symmetrical in their inner 6$^{\prime}$.
The X-ray temperatures and the bolometric luminosities are taken from
\citet{WXF}, for H$_{0}$ = 75 km s$^{-1}$ Mpc$^{-1}$, and the
HI columns are those given by \citet{DL}.

[{\it EDITOR: INSERT TABLE~\ref{t01} HERE.}]

Because galaxy clusters have radial temperature gradients, which would lead
to radial softness variation, we examined softness variations azimuthally,
either in circular annuli, or in the case of clusters with elliptical
isophotes, in elliptical annuli.  We avoided the central region and used
four annular regions with radii (or semi-major axis) of
1$^{\prime}$-2$^{\prime}$ (annulus 1), 2$^{\prime}$-3$^{\prime}$
(annulus 2), 3$^{\prime}$-4.25$^{\prime}$ (annulus 3) and
4.25$^{\prime}$-5.5$^{\prime}$ (annulus 4).  We also examined larger annuli,
but the surface brightness was generally too low to be of use.  In addition,
each annular region was divided into 24 regions of 15\dgr\ each, with region
1 being from position angle 0\dgr\ to 15\dgr, with the position angle
increasing east from north as is the usual convention.  This leads to 96
independent positions per cluster.

The signature that we are looking for in the X-ray emission is a local
hardening of the emission due to the soft photons being absorbed by an
excess in the Galactic HI.  This will lead to a decrease in the hardness
ratio with the change due to decrease in the number of soft photons rather
than an increase in the number of hard photons.  We discuss each of the
systems in turn.

\begin{figure}
\epsscale{0.8}
\plotone{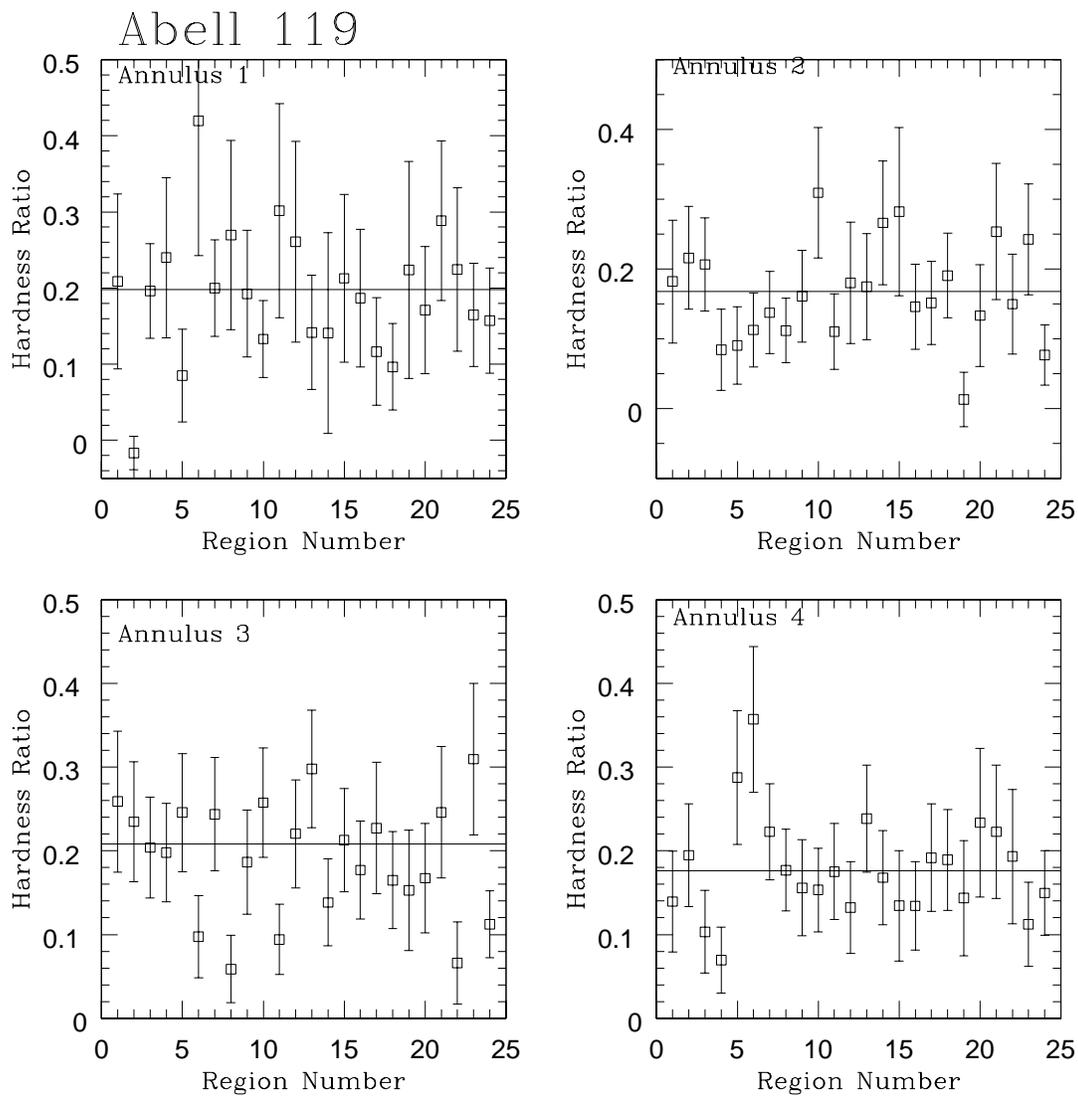}
\caption{The ratio of the fluxes in the 0.2-0.5 keV band to that in the
0.5-2.0 keV band for 15\dgr\ regions within annuli at
1$^{\prime}$-2$^{\prime}$ (annulus 1), 2$^{\prime}$-3$^{\prime}$ (annulus 2),
3$^{\prime}$-4.25$^{\prime}$ (annulus 3), and
4.25$^{\prime}$-5.5$^{\prime}$ (annulus 4),
centered on the X-ray center of Abell 119.  Several regions have hardness
ratios that deviate significantly from the mean, such as at:
annulus 1, region 2; annulus 2, region 19; and annulus 3, region 8.
Deviations of this nature can be caused by excess gas along the line of
sight, which preferentially absorbs the soft X-ray emission.
\label{f05}}
\end{figure}

Abell 119:  The hardness ratio is shown as a function of annular ring and
region (position angle) in Figure~\ref{f05}.  There are three positions for
which the hardness ratio is more than 3$\sigma $ from the mean of the sample
(annulus 1, region 2; annulus 2, region 19; annulus 3, region 8).  In each
case, the deviation is in the sense that the region is harder than the mean
and it is the soft band that has decreased relative to it neighbors rather
than an increase in the hard band.  These are precisely the expectations
for regions with excesses in their Galactic HI columns.  Also, the next
four deviant points in the sample (about 2.5$\sigma $; regions 6, 11, and 22
in annulus 3 and region 4 in annulus 4) are also harder than the average and
due to a decrease in the number of soft photons.

Abell 2029:  There is one region that deviates by more than 3$\sigma $ and
is in annulus 1 at region 10, but this is due to a positive deviation in
the hard flux rather than a decrease in the soft flux, so this does not meet
our criteria.

\begin{figure}
\epsscale{0.8}
\plotone{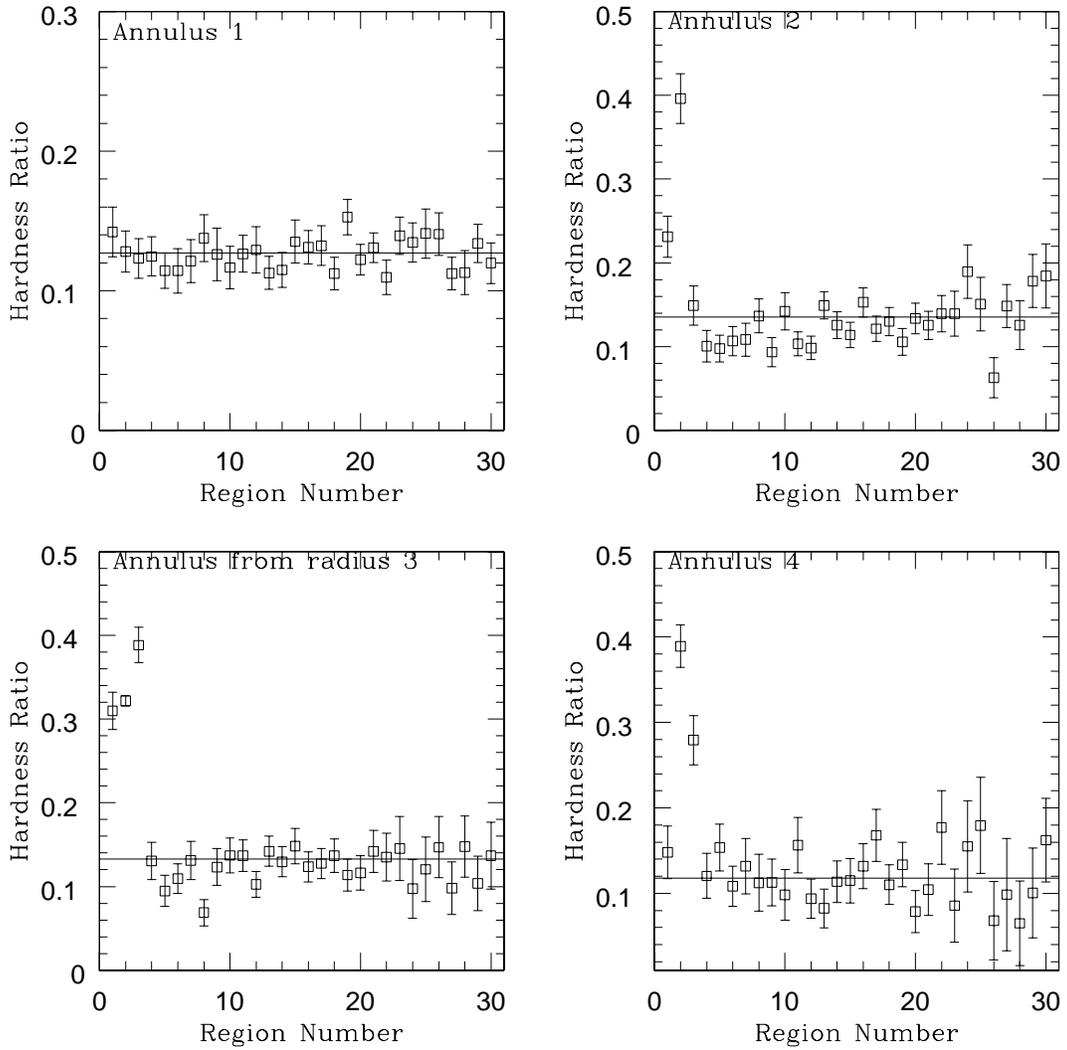}
\caption{The same ratio as in Figure 4, but for elliptical annuli centered
on the X-ray center of Abell 2142.  There is a strong source of extended
emission going from annuli 2-4 in regions 1-3, so these regions were not
used in calculating the mean hardness ratio.  There are two regions that
deviate more than 3$\sigma $ from the mean (annulus 2, region 26; annulus 3,
region 8) and they are harder, consistent with excess absorption by
foreground gas.
\label{f06}}
\end{figure}

Abell 2142:  In this system, we used elliptical isophotes with an
eccentricity of 0.82 and an orientation of the major axis of $-$58\dgr.
There is a strong point source that could not be removed and it affects
the first three regions in annuli 2, 3, and 4 (at PA 0-45\dgr),
so they were removed from the analysis.  The photon statistics were
good for this long observation, so the annuli were divided into 30 regions,
each 12\dgr\ wide.  There are two regions with deviations more than 3$\sigma$
(annulus 2, region 26; annulus 3, region 8), both regions being harder than
the average and the hardness is due to the lack of soft photons rather than
an excess of hard photons (Figure~\ref{f06}).

For the remaining two systems, AWM4 and Abell 3376, there was no evidence of
deviations in the hardness ratio beyond 3$\sigma $, although the numbers of
photons were less than the other clusters, so the statistical significance
is poorer.

To summarize, there were about 600 locations in which we could have seen
fluctuations in the hardness ratio, although about half those were of lower
statistical quality than the others.  Six locations deviated from the mean
hardness ratio in their annuli by more than 3$\sigma $, all toward the
region of fewer soft photons.  About one such deviation is expected from
statistical fluctuations, so this represents an excess.  Five of these six
regions were harder because of a paucity of soft photons, as would be
expected if there were a positive excess in the Galactic neutral hydrogen
column.  We cannot rule out that these statistically meaningful
fluctuations in the hardness ratio is intrinsic to the cluster.  To
determine whether the fluctuations are due to variations in the Galactic gas
will require an accurate HI map in this region.  In addition, we will be
able to more accurately define the hardness variations with future X-ray
observations, especially with XMM-Newton.

Although point sources are largely removed in the data processing, we
comment on the effect by point sources that are unresolved and therefore
would not be removed.  In the ROSAT band, the typical point source is of
comparable hardness to the thermal Bremsstrahlung X-ray emission from a
galaxy cluster.  Also, the surface brightness of a galaxy cluster is much
greater than that of the unresolved X-ray background sources.  Taken
together, we expect that unresolved point sources will not influence our
analysis.

\section{Applications} 

The fluctuations in the foreground absorption column can have a number of
effects, each of which should be modeled.  These fluctuations can lead to
variable absorption and reddening, which will affect the properties of an
HR\ diagram of a star cluster.  Another example occurs in absorption line
studies (as discussed above), since the mean neutral gas column inferred
from 21 cm emission studies is likely to be higher than the median line of
sight, which is more relevant for individual lines of sight.  A third
example, which we explore in more detail, is the modeling of galaxy
clusters, since this will be a systematic effect.  This systematic effect
either leads one to obtain an X-ray absorption column lower than the mean
column (which would help explain the results of \citet{AB99b}, or
alternatively, it will lead to a flux residual that might be attributed
to a new "soft" component of the X-ray emission.

The presence of a new soft X-ray component was first presented by \citet{Letal}
from data obtained with the EUVE satellite, combined with ROSAT
X-ray data.  The instrument used by the EUVE for this detection was the DS
telescope \citep{BM}, which is sensitive to photons up to
about 0.19 keV, which is the soft X-ray region.  Soft X-rays can be
absorbed by Galactic gas, and for a moderately low column density of
$2\tenup{20}$\acm, the optical depth to continuous absorption is
unity at 0.24 keV and $\tau=3$ at 0.16 keV, so the X-rays that are detected
in the EUVE band lie in the heavily absorbed region of the X-ray spectrum.
\citet{Letal} took the X-ray spectrum of some bright galaxy clusters and fit
an absorption spectrum with a single column density and then compared the
flux in the EUVE band to the measured value.  They found that the EUVE
emission was brighter than anticipated, leading them to conclude that there
was a very soft X-ray excess in the emission that needed to be explained.

\begin{figure}
\epsscale{0.8}
\plotone{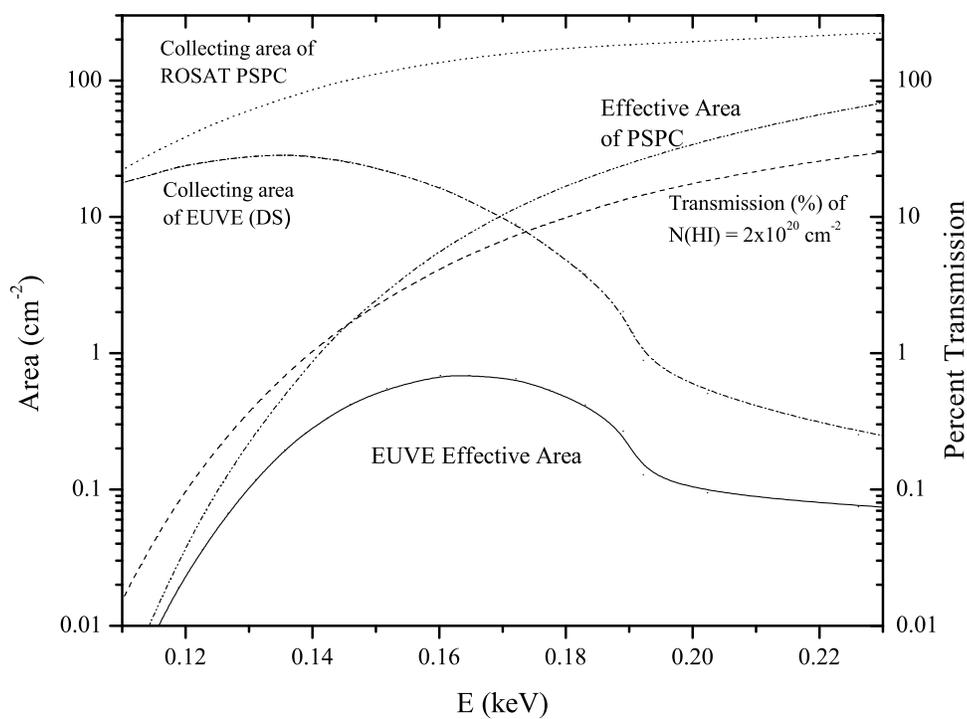}
\caption{The raw and effective collecting areas for the EUVE DS telescope
and the ROSAT PSPC in the energy range 0.11 - 0.23 keV.  The effective area
is the product of the raw collecting area and the fractional transmission
through a Galactic gas column of $2\tenup{20}$\acm.  The EUVE effective area
has a bandwidth of about 0.14-0.19 keV (FWHM), so it samples a well-defined
waveband, but it has an order of magnitude less effective collecting area than
the ROSAT PSPC.
\label{f07}}
\end{figure}

The validity of this emission has been questioned from two other works on
the subject.  \citet{AB00} pointed out that the same soft
excess should be visible in the ROSAT PSPC as well, and possibly easier to
detect.  The effective areas of the PSPC and the EUVE DS telescope are
shown in Figure~\ref{f07}, which includes not only the raw collecting area of
both instruments, but the effective collecting are due to absorption by
Galactic gas with a column density of $2\tenup{20}$\acm, which is fairly
typical of the sight lines involved.  One finds that the effective EUVE
bandpass is 0.144-0.186 keV (FWHM), which can be a great aid because
although the PSPC has more effective collecting area in this energy range,
its energy resolution is significantly poorer than the EUVE bandpass of 0.04
keV.  In this EUVE bandpass, the optical depth is about 3, so if the
Galactic column has been overestimated slightly, it makes a large difference
in the predicted flux.  Galactic columns out of the plane, such as these,
are difficult to determine accurately, and \citet{AB99b} showed that in nearly
all cases, the soft X-ray excess would vanish if one adopted a column that was
less than 3$\sigma $ from the published values.  The only exception was the
Coma cluster, for which the Galactic column is particularly small.  For low HI
columns, the corrections to the 21 cm observations are the greatest, so it is
possible that the HI column is particularly low and that there is no soft
X-ray excess.

The validity of the soft X-ray excess was also questioned by \citet{BBK},
who showed that certain data processing issues could complicate the ability to
analyze the data.  In particular, they showed that flat-fielding was a key
issue, so they constructed flat fields from many long observations, and when
this was applied, soft excesses vanished in Abell 2199 and Abell 1795, two
clusters that \citet{LBM} claimed to have excesses.  However, they find that
the Coma cluster still has an excess in its soft X-ray emission, the same
result as \citet{AB99b}, working from the ROSAT data.

\begin{figure}
\epsscale{0.8}
\plotone{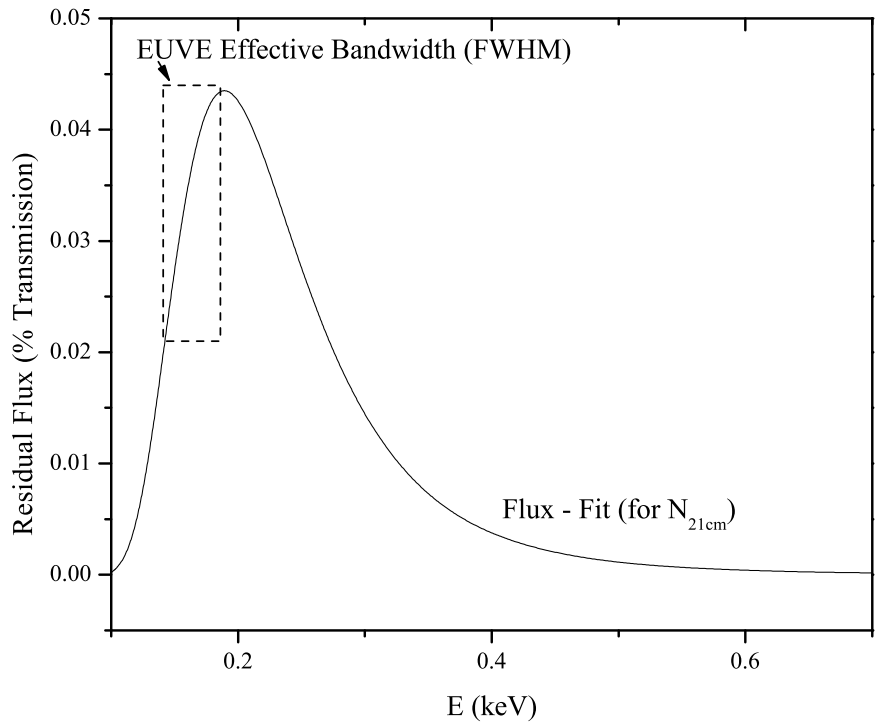}
\caption{The residual between the incident flux and the model for the
flux when using the mean column density (this is the difference between the
two lines in Figure 1).  The dashed box shows the FWHM of the EUVE DS
instrument along the energy axis.
\label{f08}}
\end{figure}

\begin{figure}
\epsscale{0.8}
\plotone{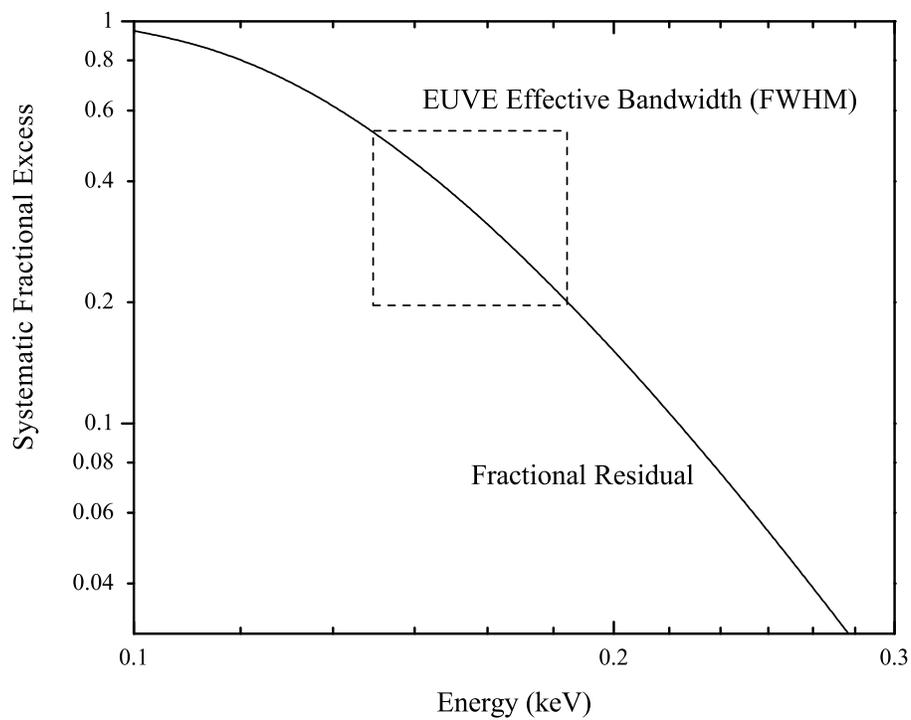}
\figcaption{The same residual shown in Figure 7, but expressed as the
ratio of the residual flux to the model flux when using the average column
density.  Within the dashed box, which shows the effective bandwidth of the
EUVE DS in energy, the residual flux can be 20-60\% of the expected flux (a
flux-weighted value of about 35\%), comparable to the excess soft X-ray
emission that has been discussed in the literature.
\label{f09}}
\end{figure}

Here we show that even if one has measured the mean Galactic 21 cm column
correctly, fluctuations in that HI column lead to a soft X-ray excess.  As
an example, when we use the model described in \S 2, the residual flux has a
peak at about 0.19 keV and a substantial amount of flux falls in the EUVE
band (Figure~\ref{f08}).  In the EUVE band, the fractional excess is claimed to
be about 20-50\% above the expected values, and that is approximately the
amount that we predict from our model of column density fluctuations
(Figure~\ref{f09}).

Ideally, one should use the detailed column density map associated with the
particular clusters in question, but 21 cm data of this sort does not exist.
Consequently, we cannot demonstrate that this effect is responsible for
the extraction of a soft X-ray excess; we merely suggest it as a possibility.

\section{Discussion and Final Comments}

We have shown the potential implications of variations in the Galactic
absorption column in trying to interpret a variety of observations.  The
greatest shortcoming in determining the true importance of this effect is
the lack of accurate data on the HI variations on the sky at small angular
scale.  Such observations need a combination of synthesis array
observations in combination with single dish data in order to include the
"zero spacing" data.  Aside from the two observations discussed above, data
of this kind have been largely confined to the Galactic plane, where the
Dominion Radio Astronomy Observatory has undertaken a survey from
$74^{\circ} < l < 147^{\circ}$, but in the narrow latitude band
$-3.5^{\circ} < b < 5.5^{\circ}$.  This is far from the region where most of
the galaxy clusters lie, typically $20^{\circ} < |b|$.

Another issue that needs to be examined is the degree to which the HI
fluctuations in these fields follows the DIRBE/IRAS maps of \citet{SFD}.  If
it could be shown that the HI generally follows the DIRBE/IRAS data to good
accuracy, then one could use that data as a proxy for the HI data.  Currently,
the situation is unclear since in the WW 187 field, there is a poor correlation
between the HI column and the DIRBE/IRAS data, while in the IRAS filament
mapped by Joncas, the correlation is much better.

The are a variety of effects that will be clarified by future HI
observations, so we strongly encourage observers to obtain such data.

\vspace{10pt}

We are especially indebted to Bart Wakker and Gilles Joncas for sending us
copies of their data for us to analyze.  Also, we would like to thank
Thomas Bergh\"{o}fer for his patience in answering our many questions about
EUVE data.  We would like to thank Jay Lockman, Jimmy Irwin, Renato Dupke,
Morton Roberts, and Stu Bowyer for their advice and comments.  Partial
support for this work has been provided by NASA through LTSA grant
NAG5-10765.

\clearpage

\clearpage

\clearpage

\clearpage

\clearpage

\clearpage

\clearpage

\clearpage

\clearpage

\clearpage

\begin{center}
\begin{deluxetable}{lccllccc}
\tablewidth{484pt}
\tablecaption{Galaxy Cluster Properties\label{t01}}
\tablehead{
\colhead{Cluster} &
\colhead{RA} &
\colhead{DEC} &
\colhead{$z$} &
\colhead{$L_x$} &
\colhead{$T$} &
\colhead{$N_{HI}$} &
\colhead{$t_{\exp }$} \\
\colhead{Name} &
\colhead{(J2000)} &
\colhead{(J2000)} &
\colhead{ } &
\colhead{($10^{44}$ erg s$^{-1}$)} &
\colhead{(keV)} &
\colhead{($10^{20}$\acm)} &
\colhead{(ksec)}
}
\startdata
Abell 119  & 00:56:16.8 & $-$01:15:00 & 0.044  &\gp3.2  & 5.6  & 3.45 & 15.2 \\
Abell 3376 & 06:00:43.6 & $-$40:03:00 & 0.0455 &\gp2.1  & 4.0  & 4.81 & 12.0 \\
Abell 2029 & 15:10:58.7 & $+$05:45:42 & 0.0767 &\gq19.  & 8.5  & 3.18 & 15.6 \\
Abell 2142 & 15:58:16.1 & $+$27:13:29 & 0.0899 &\gq27.  & 9.7  & 3.86 & 44.2 \\
AWM 4      & 16:04:55.8 & $+$23:55:54 & 0.0323 &\gp0.30 & 2.4  & 5.03 & 19.9 \\
\enddata
\end{deluxetable}
\end{center}

\end{document}